\documentclass[%
 reprint,
 amsmath,amssymb,
 aps,
]{revtex4-2}

\usepackage{graphicx}% Include figure files
\usepackage{dcolumn}% Align table columns on decimal point
\usepackage{bm}% bold math
\begin{document}

\preprint{}

\title{Gauge symmetries and the Higgs mechanism in Quantum Finance}% Force line breaks with \\

\author{Ivan Arraut}
 \altaffiliation{University of Saint Joseph\\
  Estrada Marginal da Ilha Verde, 14-17, Macao, China.}%Lines break automatically or can be forced with \\
%\affiliation{%


%\affiliation{
%\affiliation{
%\affiliation{%


\date{\today}% It is always \today, today,

\begin{abstract}
By using the Hamiltonian formulation, we demonstrate that the Merton-Garman equation emerges naturally from the Black-Scholes equation after imposing invariance (symmetry) under local (gauge) transformations over changes in the stock price. This is the case because imposing gauge symmetry implies the appearance of an additional field, which corresponds to the stochastic volatility. The gauge symmetry then imposes some constraints over the free-parameters of the Merton-Garman Hamiltonian. Finally, we analyze how the stochastic volatility gets massive dynamically via Higgs mechanism.
\end{abstract}

\maketitle

%\tableofcontents

\section{Introduction}

The dynamic of any system can be always analyzed by using the Lagrangian formulation or the Hamiltonian one \cite{1, 2, 3}. Selecting one or the other formulation is a matter of convenience. The Lagrangian formulation for example, is the favorite one in particle physics because in such a case, relativistic effects emerge when the particles accelerate at velocities near to the speed of light \cite{1}. The Lagrangian, being an invariant under Lorentz transformations, is then the ideal mathematical object for analyzing relativistic situations. The Hamiltonian formulation, on the other hand, is the favorite one when we do not deal with relativistic effects because the Hamiltonian itself is not a relativistic invariant \cite{4, 5, 6}. Both formulations, Lagrangian and Hamiltonian, contain the same amount of information and they are connected to each other through the Legendre transformation \cite{1,2,3, 7}. No matter which formulation we select, in both cases we always have a kinetic term plus a potential term. When we want to analyze the equilibrium condition of a system (vacuum condition) by using the Hamiltonian formulation, we then ignore the kinetic terms and we then proceed to find the extreme conditions for the potential term. This general statement can be verified in other research areas, including gravity \cite{Mypapers1, Mypapers11, Mypapers12, Mypapers13, Mypapers14, Mypapers15}, condensed matter physics \cite{Mypapers, Mypapers31, Mypapers32}, in (Quantum) finance \cite{Mypapers2, Mypapers21} as well as in Quantum Field Theory in general \cite{1}. In ordinary circumstances, there is a single vacuum state or equilibrium condition for the system. However, the system has a set of free-parameters and for certain values taken by these parameters, the vacuum configuration might change in a way that it becomes degenerate (multiplicity of ground states or martingale states if we consider finance) \cite{1, Mypapers1, Mypapers11, Mypapers12, Mypapers13, Mypapers14, Mypapers15, Mypapers, Mypapers31, Mypapers32, Mypapers2, Mypapers21, Ground, Ground2, Ground3, Ground4}. This means that under such conditions, there are multiple equivalent (but disconnected) ground states and the system at the end selects randomly one of them. When this occurs, we say that one of the symmetries of the system is spontaneously broken \cite{Spo, Spo1, Spo2, Spo3, Spo4}. Spontaneous symmetry breaking is a term used for describing symmetries satisfied by the Hamiltonian but violated by the ground state. Spontaneous symmetry breaking might involve global or local (gauge) symmetries \cite{gauge}. In both cases, gapless particles dubbed Nambu-Goldstone bosons emerge \cite{Mypapers, Mypapers31, Mypapers32, Spo, Spo1, Spo2, Spo3}. When we deal with local (gauge) symmetries, once the system selects its ground state among all the possibilities, then the Nambu-Goldstone bosons are eaten up by another field, which becomes massive in this way \cite{Higgs, Higgs2}. Inside the scenario of Quantum finance, applied to the stock market prediction, the Hamiltonian formulation has been developed \cite{Baquie}. By using this formalism, the spontaneous symmetry breaking phenomena in Quantum Finance, has been analyzed \cite{Mypapers2}. However, by the date, the local (gauge) symmetries inside the stock market equations have not been studied. Furthermore, the Higgs mechanism in Quantum Finance has not been analyzed yet. In this paper, we demonstrate that the Merton-Garman (MG) equation emerges naturally from the Black-Scholes (BS) equation after imposing the gauge symmetry condition under changes on the prices of the stock over the BS equation. This surprising result means that the BS equation is locally equivalent to the MG equation, although globally different. Analogue results emerge from different gauge theories in physics \cite{1, Spo2, Higgs, Higgs2, Gauge2, Gauge3, Gauge4, Gauge5, Gauge6}. This demonstrates that the connection between symmetries is general and it can be extended to other research areas. Finally, we analyze the Higgs mechanism in this scenario, finding then that the stochastic volatility is the field able to become massive dynamically.

\section{The Hamiltonian formulation in the stock market}

From previous analysis, it is known that the dynamic behind the Black-Scholes equation as well as the Merton-Garman one, can be analyzed through the corresponding Hamiltonians \cite{ Mypapers2, Baquie}. Any Hamiltonian equation has the following functional form

\begin{equation}   \label{eq.1}
H \psi(S, t)=E\psi(S, t).    
\end{equation}
For the BS case, the Hamiltonian is given by the equation

\begin{equation}   \label{BSHamiltonian}
\hat{H}_{BS}=-\frac{\sigma^2}{2}\frac{\partial^2}{\partial x^2}+\left(\frac{1}{2}\sigma^2-r\right)\frac{\partial}{\partial x}+r,
\end{equation}
where
\begin{equation}   \label{StockPrice}
S=e^x,\;\;\;\;\;-\infty\leq x\leq\infty. 
\end{equation}
Here $S(t)$ is the price of the stock. In eq. (\ref{BSHamiltonian}), $\sigma$ is the volatility and $r$ is the interest rate. While the interest rate is easy to fix, the volatility is normally difficult to estimate. In fact, different persons could estimate different values for the volatility of the same Option under the same conditions \cite{Volatility}. No matter what is the estimated value for the volatility, the dynamic of the BS system is determined by the relation between $\sigma$ and $r$. The function $\psi(S, t)$ appearing in eq. (\ref{eq.1}), corresponds to the price of an Option. The Option could be a Put Option or a Call Option, depending on the type of contract generated between the holder of the Option and the writer of the same Option \cite{Volatility, Vol2}. No matter what type of contract the parties agree, the Black-Scholes equation, here expressed in its Hamiltonian form, is able to find the fair price of an Option as a function of the stock price and the time \cite{Volatility}. The significant advantage of the Hamiltonian formulation is that even without solving the equations explicitly, still we can understand the dynamic of the system by analyzing the behavior of the vacuum (ground) state, which inside this scenario, appears as the Martingale state. The Martingale state in the stock market is defined as the state annihilated by the BS Hamiltonian. Then the Martingale conditions can be expressed as \cite{Baquie}
\begin{equation}   \label{eq.2}
H_{BS}\vert S>=0.    
\end{equation}
Here the ket $\vert S>$ is identified as the Martingale state.

\section{General definition of a martingale state}

When the market is complete and the no-arbitrage condition holds, then we can claim the existence of the martingale state. The martingale condition corresponds to a risk-neutral measure such that the conditional probability of the discounted value of an equity at time $t\neq0$ is just equal to its present value at $t=0$ \cite{Baquie, Martingale, Martingale2, Martingale3, Martingale4, Martingale5}. This can be expressed Mathematically as 

\begin{equation}   \label{martingale}
E[X_{n+1}\vert x_1, x_2, ..., x_n]=x_n,    
\end{equation}
as a general statement for any process. For the specific case of the stock prices $S(t)$, the martingale state can be expressed as 

\begin{equation}   \label{interpret}
S(0)=E\left[e^{-\int_0^tr(t')dt'}S(t)\vert S(0)\right].
\end{equation}
If we define the martingale state as $S=e^x$, and if we take the interest rate as a constant $r$, then it is possible to demonstrate that

\begin{equation}
\vert S>=e^{-(t_*-t)\hat{H}}\vert S>.    
\end{equation}
This result is just equivalent to eq. (\ref{eq.2}). In ordinary cases, the martingale state is unique. However, in the BS equation, the possibility of having spontaneous symmetry breaking and then a multiplicity of martingale states has been studied before \cite{Mypapers2}. Independent on how many ground states (equilibrium conditions) the system has, the market equilibrium condition is reached when all the participants in the market share the same amount of information. In these situations there is no winner nor loser until some random fluctuations break the market equilibrium.

\section{Spontaneous symmetry breaking in Quantum Finance}

Spontaneous symmetry breaking was explored before inside the scenario of Quantum Finance for the cases involving some global symmetries \cite{Mypapers2}. For the Hamiltonian defined in eq. (\ref{BSHamiltonian}), it has been proved before that the symmetry under changes in prices is spontaneously broken. Then the relation

\begin{equation}   \label{beokensym}
\hat{p}\vert S>\neq0,
\end{equation}
is valid. The operator $\hat{p}$ maps one martingale state $\vert S>$ into another one $\vert S'>$. This issue can be perceived better when if we consider the formal definition of the operator $\hat{p}$ by looking on its action over some ket $\vert S>$, which is defined as

\begin{equation}   \label{Derivative}
<x\vert\hat{p}\vert S>=\frac{\partial}{\partial x}C(x, t),    
\end{equation}
with $C(x, t)=<x\vert S>$. The derivative operation in eq. (\ref{Derivative}), generates a shift on the ground state, mapping then the equilibrium condition toward another one. The broken symmetry defined in eq. (\ref{beokensym}), can be considered to be a local symmetry. In \cite{Mypapers2}, a rigorous analysis of the cases involving local symmetries was not done. It is important to note that although different martingale states are connected by a broken generator, they do not have necessarily the same stock price associated to them. Still, all the different ground states represent equilibrium conditions in the market. It is evident that the operator $\hat{p}$, commutes with the BS Hamiltonian in eq. (\ref{BSHamiltonian}) and then it represents a symmetry for the system.

\subsection{Imposing gauge symmetries inside the Black-Scholes Hamiltonia: The origen of the Merton-Garman equation}

We can now analyze the behavior of the BS Hamiltonian under local transformations involving changes of the prices of the stock. We can take a local transformation under the changes of prices as $U=e^{\omega \theta(x)}$. Here $\theta(x)$ is a variable depending on $x$, which also depends on the price of the stock $S$ as the eq. (\ref{StockPrice}) suggests. If the operator $U$ were a symmetry of the system, then it would satisfy the condition $[\hat{H}_{BS}, U]=0$. However, it is possible to demonstrate that this is not the case here after using the Hamiltonian given in eq. (\ref{BSHamiltonian}). In fact, if we use eq. (\ref{BSHamiltonian}), together with the definition of the local changes in price, then we have

\begin{equation}   \label{NESym}
[\hat{H}_{BS}, U]\neq0.    
\end{equation}
After some calculation, it is possible to demonstrate that in order to get an exact symmetry under local changes of the prices ($U=e^{\omega\theta(x)}$), then the BS Hamiltonian needs to add certain terms inside its definition in eq. (\ref{BSHamiltonian}). Under the action of the local transformation $U=e^{\omega\theta(x)}$, the BS Hamiltonian is changed as
\begin{eqnarray}   \label{ModHamiltonia}
\hat{H}_{BS}\to \hat{H}_{BS}+\frac{\sigma^2\omega(1+\omega)}{2}\left(\frac{\partial\theta(x)}{\partial x}\right)^2+\nonumber\\
\sigma^2\omega\left(\frac{\partial\theta(x)}{\partial x}\right)\frac{\partial}{\partial x}+\omega\left(\frac{1}{2}\sigma^2-r\right)\frac{\partial\theta(x)}{\partial x}.
\end{eqnarray}
Then the BS equation does not satisfy the gauge symmetry under changes of prices defined through the transformation $U=e^{\omega\theta(x)}$. For restoring the gauge-invariance, we have to extend the standard derivative in eq. (\ref{BSHamiltonian}), such that it becomes a covariant derivative. Without loss of generality, here we will define the covariant derivative as

\begin{equation}   \label{CovDer}
\frac{\partial}{\partial x}\to \frac{\partial}{\partial x}+\hat{p}_y.  
\end{equation}
Here we interpret $\hat{p}_y$ as the momentum associated with the stochastic volatility. After replacing the ordinary derivative with the covariant derivative in eq. (\ref{BSHamiltonian}), we obtain
\begin{eqnarray}   \label{MertonMOdified}
\hat{H}_{BS}\to\hat{H}=\frac{\sigma^2}{2}\left(-\hat{p}_x-\hat{p}_y\right)\left(\hat{p}_x+\hat{p}_y\right)\nonumber\\
+\left(\frac{1}{2}\sigma^2-r\right)\left(\hat{p}_x+\hat{p}_y\right)+r.
\end{eqnarray}
The minus sign difference in the first term appears because the momentum associated to the changes of prices, as well as the momentum associated to the changes on the stochastic volatility are both non-Hermitian quantities, satisfying then the conditions 

\begin{equation}
\hat{p}_x^+=\frac{\partial}{\partial x}^+=-\frac{\partial}{\partial x}, \;\;\;\hat{p}_y^+=\frac{\partial}{\partial y}^+=-\frac{\partial}{\partial y}.
\end{equation}
Here the index $+$ means Hermitian conjugate operation. After an expansion, the equation (\ref{MertonMOdified}) becomes

\begin{eqnarray}   \label{MertonMOdified2}
\hat{H}=-\frac{\sigma^2}{2}\hat{p}_x^2+\left(\frac{1}{2}\sigma^2-r\right)\hat{p}_x-\frac{\sigma^2}{2}\hat{p}_y^2-\sigma^2\hat{p}_x\hat{p}_y\nonumber\\
+\left(\frac{1}{2}\sigma^2-r\right)\hat{p}_y+r.
\end{eqnarray}
The gauge invariance under a general transformation of the form $U=e^{\omega\theta(x, y)}$ for the new financial Hamiltonian defined in eq. (\ref{MertonMOdified2}) is guaranteed if the following conditions are satisfied
\begin{eqnarray}
\left(\frac{\partial\theta}{\partial x}\right)^2=\frac{\omega}{1+\omega}\left(\frac{\partial\theta}{\partial y}\right)^2,\nonumber\\
\left(\frac{\partial\theta}{\partial x}\right)\hat{p}_x=\left(\frac{\partial\theta}{\partial y}\right)\hat{p}_y,\nonumber\\
\frac{\partial\theta}{\partial x}+\frac{\partial\theta}{\partial y}-4\frac{\partial^2\theta}{\partial x\partial y}=\frac{2r}{\sigma^2}\left(\frac{\partial\theta}{\partial x}+\frac{\partial\theta}{\partial y}\right).
\end{eqnarray}
These conditions are obtained after checking the invariance of eq. (\ref{MertonMOdified2}). In this way the changes due to local transformations of the new terms appearing in eq. (\ref{MertonMOdified2}), cancel exactly the additional terms appearing in eq. (\ref{ModHamiltonia}). Note that interestingly when $\sigma^2=2r$, then $\frac{\partial^2\theta}{\partial x\partial y}=0$. This condition corresponds, additionally, to the Hermiticity condition for the BS Hamiltonian. Independent of the values taken by the free-parameters of the system, as far as we can obtain the correct function $\theta(x, y)$, the gauge invariance of the Hamiltonian (\ref{MertonMOdified2}) is satisfied. The Hamiltonian defined in eq. (\ref{MertonMOdified2}) is the Merton-Garman Hamiltonian if we redefine the parameters appropriately as follows 

\begin{eqnarray}   \label{Volcoeff}
\zeta^2=e^{-2y\left(\alpha-\frac{3}{2}\right)},\nonumber\\
\rho\zeta=e^{-y\left(\alpha-\frac{3}{2}\right)},\nonumber\\
r=\lambda e^{-y}+\mu.
\end{eqnarray}
These expressions guarantee the equivalence of the Hamiltonian in eq. (\ref{MertonMOdified2}) and the MG Hamiltonian \cite{Baquie}. It is interesting to notice that the relations in eq. (\ref{Volcoeff}) give us the conditions $\rho=\pm1$, which are the extreme conditions for the parameter $\rho$. In the standard analysis of the MG equation, the parameter $\rho$ respects the following condition

\begin{equation}
-1\leq\rho\leq1.    
\end{equation}
Then for the BS and the MG equations to be connected through gauge invariance under changes of the prices of the stock market system, as far as we define the covariant derivative as in eq. (\ref{CovDer}), then the MG parameter $\rho$ can only take the extreme values. In this way, the white noises related to the time evolution of the stock price and volatility, satisfy the following conditions
\begin{equation}
<R_1(t)R_1(t')>=<R_2(t)R_2(t')>=\pm<R_1(t)R_2(t')>,    
\end{equation}
when the gauge invariance connects the BS and the MG equations \cite{Baquie, Mypapers21}. Finally, we must remark the interesting connection between the interest rate $r$ and the volatility coefficients $\lambda$ and $\mu$ inside eq. (\ref{Volcoeff}). In this way, when the BS and the MG equations are connected through the local symmetry transformations, the interest rate and the volatility are related through the parameters $\lambda$ and $\mu$.

\section{The Higgs mechanism in Quantum Finance: The dynamical origin of the volatility}

It is not a surprise to claim that the volatility behaves as a massive term inside the financial Hamiltonian. The Hamiltonian obtained in eq. (\ref{MertonMOdified2}) is the Merton-Garman Hamiltonian as far as the conditions (\ref{Volcoeff}) are satisfied. Then we can safely suggest that the vacuum condition for the Hamiltonian in eq. (\ref{MertonMOdified2}) is given by 

\begin{equation}   \label{newmartin}
\hat{H}_{MG}e^{x+y}=\hat{H}_{MG}S(x, y, t)=0.    
\end{equation}
We can consider this as the martingale condition for the Hamiltonian (\ref{MertonMOdified2}). In \cite{Mypapers2}, it was proved that the martingale condition (\ref{newmartin}) is a real vacuum condition when the following constraint 
\begin{equation}   \label{corona4}
\lambda+e^y\left(\mu+\frac{\zeta^2}{2}e^{2y(\alpha-1)}+\rho\zeta e^{y(\alpha-1/2)}\right)=0,
\end{equation}
is satisfied. If we impose the constraints defined in eq. (\ref{Volcoeff}), inside eq. (\ref{corona4}), then we get 

\begin{equation}   \label{THisonemama}
e^{2y}+\mu e^y+\lambda=0.    
\end{equation}
If we solve this equation, we get

\begin{equation}
e^y=-\frac{\mu}{2}\left(1\mp\sqrt{1-\frac{4\lambda}{\mu}}\right).    
\end{equation}
This equation suggests a relation between the stochastic volatility and the parameters $\lambda$ and $\mu$. This in addition implies that the martingale condition (\ref{newmartin}), is well-defined when there is a direct relation between the volatility and the interest rate. This can be seen explicitly if we introduce the last relation in eq. (\ref{Volcoeff}) inside eq. (\ref{THisonemama}). In such a case, we get $e^y=-r$. This condition, then implies a negative interest rate in order to have a market equilibrium. The martingale state suggested in eq. (\ref{newmartin}) is not the only possible definition of a martingales state. However, it is the only definition involving the prices of the stock as well as the stochastic volatility. No matter which martingale definition we use, we can analyze the general equilibrium in the market when we look into the potential term of the MG Hamiltonian defined in eq. (\ref{MertonMOdified2}). Then we can define the potential as \cite{Mypapers2, Mypapers21}

\begin{equation}   \label{OperatoirC}
\hat{V}=\left(\frac{1}{2}\sigma^2-r\right)\hat{p}_x-\sigma^2\hat{p}_x\hat{p}_y+\left(\frac{1}{2}\sigma^2-r\right)\hat{p}_y+r.
\end{equation}
Here we consider the terms linear in $\hat{p}_x$ and $\hat{p}_y$ as potential terms. We can compare this eq. with the potential term for the MG Hamiltonian, analyzed in \cite{Mypapers2} and repeated here as

\begin{eqnarray}   \label{happotential}
\hat{V}(x, y)= -\left(r-\frac{e^y}{2}\right)\hat{p}_x-\nonumber\\
\left(\lambda e^{-y}+\mu-\frac{\zeta^2}{2}e^{2y(\alpha-1)}\right)\hat{p}_y+r.
\end{eqnarray}
Note that the equations (\ref{OperatoirC}) and (\ref{happotential}) are the same under the conditions (\ref{Volcoeff}). In the neighborhood of the minimal defined by the condition (\ref{newmartin}), we can considering that $<x, y\vert S>=S(x, y, t)=e^{x+y}=\sum_{n=0}^\infty(x+y)^n/n!=\sum_{n=0}^\infty\phi^n_x\phi^n_y$. Additionally, we know that $\partial S(x, y, t)/\partial x=\partial S(x, y, t)/\partial y=\sum_n(x+y)^{n-1}/(n-1)!=e^{x+y}=\sum_nn\phi_x^{n-1}\phi_y^n=\sum_nn\phi_x^n\phi_y^{n-1}$. Then without loss of generality, limiting the series expansion to second order, then the potential term to analyze is

\begin{eqnarray}   \label{happo2}
<x, y\vert \hat{V}(x, y)\vert S>=V(S)=-2\left(r-\frac{e^y}{2}\right)\phi_x\phi_y^2\nonumber\\
-2\left(\lambda e^{-y}+\mu-\frac{\zeta^2}{2}e^{2y(\alpha-1)}\right)\phi_x^2\phi_y+r\phi_x^2\phi_y^2.
\end{eqnarray}
Note that the conditions defined in eq. (\ref{Volcoeff}), makes the equations (\ref{happotential}) and (\ref{happo2}) to be equivalent. This means that 

\begin{equation}   \label{Rich}
\lambda e^{-y}+\mu-\frac{\zeta^2}{2}e^{2y(\alpha-1)}=r-\frac{e^y}{2},    
\end{equation}
for this case. This suggests that the coefficients involving $\phi_x\phi_y^2$ and $\phi_x^2\phi_y$ are the same under the conditions (\ref{Volcoeff}). This is a consequence of the symmetric character of the covariant derivative selected in eq. (\ref{CovDer}). If instead of the definition (\ref{CovDer}), we have used 

\begin{equation}   \label{Cagados}
\frac{\partial}{\partial x}\to \frac{\partial}{\partial x}+\gamma\hat{p}_y,      
\end{equation}
introducing then another parameter $\gamma$. In this way, the conditions (\ref{Volcoeff}) would be modified and then the coefficients involving $\phi_x\phi_y^2$ and $\phi_x^2\phi_y$ would be different if we introduce this new parameter. Then introduction of $\gamma$ in general modifies the possible values taken by the parameters of the MG Hamiltonian. Independent of the value taken by $\gamma$ in eq. (\ref{Cagados}), at the most general and fundamental level, the MG equation emerges from the BS equation after imposing local symmetry under changes of the prices over the BS equation. In this way, the basic calculations done on the previous sections are valid in general after considering the corrections due to the parameter $\gamma$ in eq. (\ref{Cagados}). However, inside this paper, we have focused on the case $\gamma=1$ for the sake of simplicity.

\subsection{The dynamical origin of the volatility}

In order to analyze the dynamical origin of the volatility, we have to analyze the vacuum or ground state of the system. In \cite{Mypapers2}, the general vacuum condition suggested a relation of the form

\begin{equation}   \label{trivacc}
\phi_{y vac}=\left(\frac{\lambda e^{-y}+\mu-\frac{\zeta^2}{2}e^{2y(\alpha-1)}}{r-\frac{e^y}{2}}\right)\phi_{x vac},    
\end{equation}
which under the condition $\gamma=1$ in eq. (\ref{Cagados}), or equivalently, under the condition (\ref{Rich}), gives the result $\phi_{x vac}=\phi_{y vac}$. The result (\ref{trivacc}) is based on the general martingale state definition given in eq. (\ref{newmartin}) \cite{Mypapers2}. Although we could in principle work around the vacuum definition given in eq. (\ref{trivacc}), the appearance of the volatility inside the ground state definition, would make it difficult to visualize the mechanism behind the dynamical origin of the mass for the volatility. Then in this section, instead of considering the martingale state as a function of price ($x$) and volatility ($y$), we consider it as a function of the price only. In this way, if we modify the step from eq. (\ref{happotential}) to eq. (\ref{happo2}) after considering the standard definition of martingale (without the volatility), then  we get

\begin{equation}   \label{happo3}
<x\vert \hat{V}(x, y)\vert S>=V(S)=-2\left(r-\frac{e^y}{2}\right)\phi_x\phi_y^2\nonumber\\
+r\phi_x^2\phi_y^2,
\end{equation}
which ignores the term $<x\vert\hat{p}_y\vert S>=\partial S(x, t)/\partial y=0$ since in this special case, we are taking $S(x, t)$ (martingale state) as a state independent of $y$. Eq. (\ref{happo3}) gives us the ordinary Martingale condition which is the same for the BS and MG cases. The ground state in eq. (\ref{happo3}) is obtained from the condition $\partial V/\partial \phi_x=0$, obtaining then \cite{Mypapers2}

\begin{equation}   \label{nonunique}
\phi_{vac}=1-\frac{\sigma^2}{2r}.     
\end{equation}
Then the field $\phi_x$ can be expanded around this ground state as

\begin{equation}   \label{Vacredef}
\phi(x)=\phi_{vac}+\bar{\phi}(x).  
\end{equation}
For understanding the effect of this field redefinition, we need to introduce the result (\ref{Vacredef}) inside eq. (\ref{happo3}). In this way, we get 

\begin{eqnarray}   \label{happo4}
<x\vert \hat{V}(x, y)\vert S>=V(S)=-2\left(r-\frac{e^y}{2}\right)(\phi_{vac}+\nonumber\\
\bar{\phi}(x))\phi_y^2
+r(\phi_{vac}+\bar{\phi}(x))^2\phi_y^2.
\end{eqnarray}
From this expression, we obtain some terms of the form $\phi_{vac}\phi_y^2$ which represent the dynamical origin of the mass of the volatility field $\phi_y$. More explicitly, the expression (\ref{happo4}) becomes
\begin{eqnarray}   \label{happo5}
<x\vert \hat{V}(x, y)\vert S>=V(S)=\nonumber\\
\left(-2\left(r-\frac{e^y}{2}\right)
+r\phi_{vac}\right)\phi_{vac}\phi_y^2+....
\end{eqnarray}

Naturally, if $\phi_{vac}$ vanishes, then the massive term corresponding to the volatility field vanishes. This demonstrates that the dynamical origin of the volatility mass emerges from the relation between the parameters $\sigma$ and $r$ in eq. (\ref{nonunique}). Since in the MG equation $\sigma^2=e^y$, then the phenomena is even more interesting than in standard situations because it involves non-linearities. Then the volatility field, generates its own mass dynamically because its influence appears inside the definition of the vacuum state $\phi_{vac}$. Finally, since the terms in eq. (\ref{happo5}) correspond to the second-order terms of the expansion of the price and volatility fields as it was done previously in \cite{Mypapers2}, then the kinetic terms in eq. (\ref{MertonMOdified2}), do not have contributions to the dynamical origin of the volatility mass, at least not at second-order. If we consider higher-order terms in the expansion, still the same arguments used for obtaining the result (\ref{happo5}) work. The behavior of the kinetic terms for the MG equation, at all orders, was analyzed in \cite{Mypapers21}. 

\section{Conclusions}

In this paper we have demonstrated that the Merton-Garman equation emerges naturally from the Black-Scholes equation when we impose local symmetry under changes of the stock prices over the Black Scholes equation. Additionally, we have elaborated the formalism for understanding the dynamical origin of the volatility mass, which is a process analogue to the Higgs mechanism in particle physics \cite{Higgs}.    \\\\

%{\bf Acknowledgements}\\

%\begin{acknowledgments}
%The author would like to thank Prof. Izumi Tsutsui for his kind attention during our discussions about these results during the last three years at the KEK High Energy Accelerator Research organization (Theory Center).
%\end{acknowledgments}

%\nocite{*}


\end{document}